Calculating power by bootstrap, with an application to cluster-randomized trials

Ken Kleinman[1], Susan S. Huang[2]

1 Corresponding author, ken.kleinman@gmail.com, Department of Population Medicine, Harvard Medical School and Harvard Pilgrim Health Care, Boston, MA, USA
2 Division of Infectious Diseases and Health Policy Research Institute, University of California Irvine School of Medicine, Irvine, CA, USA

Abstract

Background

A key requirement for a useful power calculation is that the calculation mimic the data analysis that will be performed on the actual data, once it is observed. Close approximations may be difficult to achieve using analytic solutions, however, and thus Monte Carlo approaches, including both simulation and bootstrap resampling, are often attractive. One setting in which this is particularly true is cluster-randomized trial designs. However, Monte Carlo approaches are useful in many additional settings as well. Calculating power for cluster-randomized trials using analytic or simulation-based methods is frequently unsatisfactory due to the complexity of the data analysis methods to be employed and to the sparseness of data to inform the choice of important parameters in these methods.

Methods

We propose that among Monte Carlo methods, bootstrap approaches are most likely to generate data similar to the observed data. Means of implementation are described.

Results

We demonstrate bootstrap power calculation for a cluster-randomized trial with a survival outcome and a baseline observation period.

Conclusions

Bootstrap power calculation is a natural application of resampling methods. It provides a relatively simple solution to power calculation that is likely to more accurate than analytic solutions or simulation-based calculations. It has several important strengths. Notably, it is simple to achieve great fidelity to the proposed data analysis method and there is no requirement for parameter estimates, or estimates of their variability, from outside settings. So, for example, for cluster-randomized trials, power calculations need not depend on intracluster correlation coefficient estimates from outside studies. We are not aware of bootstrap power calculation being previously proposed or explored for cluster-randomized trials, but it can also be applied for other study designs. We demonstrated power calculations for a time-to-event outcome in a cluster randomized trial setting, for which we are unaware of an analytic alternative. The method can easily be used when preliminary data is available, as is likely to be the case when research is performed in health delivery systems or other settings where electronic medical records can be obtained.

Background

Statistical power is defined as the probability of rejecting the null hypothesis, given that some particular alternative hypothesis ("the alternative") is true. Power is particularly important from the perspectives of ethics and of allocating scarce resources. It is often ethically unjustifiable to randomize more subjects than are required to yield sufficient power, and it is a waste of resources to invest time or money in studies which have little chance of rejecting the null or when power is far greater than necessary.

In many settings, the question of how to calculate power is reasonably well addressed by closed-form equations or easily tractable mathematical methods. For instance, the power for an ordinary least squares regression is described in basic textbooks [1]. Power for logistic regression can use iterative techniques or relatively simple formulae [2,3]. Major statistical packages such as SAS (SAS Institute, Cary NC) contain routines for power calculation, and both functions and packages for power calculation are available for the free and open-source R environment [4]. There are also several stand-alone packages that simplify the calculation of power, for example, PASS (NCSS Inc., Kayesville, UT).

However, there are many settings in which these simple solutions are unsatisfactory. In order for power calculations to usefully inform our planning, the methods used must conform reasonably well to the planned analysis. If we plan to study a confounded relationship using a linear regression, the power assessment must include the confounder. If we know the outcome-predictor relationship is heteroscedastic, we should not use closed-form solutions that depend on homoscedasticity. If our study design includes a baseline period, we should not use a post-only comparison for estimating the power.

One setting in which power assessment is not simple is cluster-randomized trials. In this design, a relatively small number of administrative clusters, such as hospitals, classrooms, or physician practices, are recruited. Each cluster may contain a large numbers of individuals upon whom outcomes will be measured. Rather than randomize subjects individually to treatment arms, all of the individuals within a cluster are randomized to the same treatment arm, and in practice we say that the cluster itself is randomized to one treatment arm or another

This study design often reduces cost considerably, and in many settings it is the only way to get estimates of *pragmatic* effects-- the effects of an intervention in a typical clinical population and in settings like those which non-trial patients are likely to encounter. For example, interventions on doctors to affect prescribing practices could hardly generate generalizable results if we randomize patients. We must randomize doctors, but examine the impact on patients.

Randomization by cluster leads to complications in data analysis that have long been recognized by statisticians [5,6]. This is due to the likelihood of patients within a cluster to resemble each other, or, more formally, a lack of independence between subjects. This can be parameterized as the covariance or correlation between subjects within a cluster (the intracluster correlation coefficient or ICC) or as the variance of cluster-specific parameters ($\sigma_b^2$). Valid approached include calculating summary statistics by cluster in a first step and then comparing cluster summaries by treatment arm in a second step, and mixed effects models that incorporate all individual observations in a single model [5,6].

There are several existing analytic approaches to calculating the power for cluster-randomized trials. Many of these rely on the "design effect", $1 + (m-1)\rho$, where $m$ is the number of observations per cluster and $\rho$ is the ICC [5-8]. The "effective sample size" is calculated by dividing the actual number of subjects by the design effect. Power assessment can then continue using methods for uncorrelated data, based on the effective sample size. While this approach can be surprisingly accurate, we do not recommend using it in practice. We mention the approach here because it helps clarify the importance of the ICC: with as few as 1000 subjects per cluster, increasing the ICC from 0.001 to 0.002 results in a 33% loss of effective sample size. In contrast, the confidence limits for estimated ICC are likely to be much broader than 0.001. Cluster sizes of 1000 or greater are common in trials involving health delivery systems or communities [9,10].

While the effective sample size approach is an approximation, accurate analytical approaches also depend on the design effect, and are similarly dramatically affected by the ICC. However, many approaches based on the design effect require that each cluster has an equal number of subjects, which may well not be the case. Several investigations into the impact of this have been performed, though their results are not general [11-14]. Approximate methods of incorporating the impact of variable cluster size have been proposed, however [15-17].

These analytic and approximate options for power assessment become difficult or untenable when more complex study designs are used. For example, it is often possible to record a baseline period in which neither the treatment clusters nor the control clusters receive the intervention followed by an intervention period in which only clusters so randomized receive the intervention. This design is much stronger than an "intervention period only" design, since it can account for some pre-existing or baseline differences among the clusters. Power calculation via analytic methods are known for normal-distributed outcomes in this design, (see, e.g., Murray pages 368-369 [6], Teerenstra et al.[18]). A Stata add-on due to Hemming and Marsh provides approximate power and sample size estimation with variable cluster size and can accommodate a baseline observation period [19]. For more complex designs and, e.g., dichotomous, count, or survival outcomes, analytic results may be unknown.

Another option useful in any difficult setting and in cluster randomized trials in particular is to use simulation, as follows. First, generate data resembling the data anticipated for the study under the specific alternative hypothesis for which a power estimate is required, then perform the planned test on that data. Repeat this process many times: the proportion of the simulated data sets in which the null hypothesis is rejected is an estimate of the power. This approach is very powerful, and has been implemented for cluster-randomized trials with baseline observation periods in at least one package for R [20,21]. The package also accommodates more general crossover trials.

But despite the robustness of simulation-based methods to some design issues, they share one key weakness with the analytic approach: it is often extremely difficult to obtain credible estimates of the ICC or $\sigma_b^2$. Assessments of the variability of the ICC or $\sigma_b^2$ are even harder to find, and small differences in these parameters can lead to large differences in the estimated power, as was demonstrated using the effective sample size approximation. The difficulty of obtaining estimates has led to reliance on rules of thumb and to articles which report ranges of ICCs, to serve as reference [22]. While perhaps better than no estimate at all, estimates from unrelated areas may lead to poor estimates of power.

In addition, covariate imbalance between arms is likely when few units are randomized. Though there remains debate among trialists about whether covariate adjustment is ever appropriate, it may be thought desirable in the case of a cluster-randomized trial. If so, the adjustment should also be incorporated into the power assessment. As the model gets more complex and parameters multiply, we should have less confidence in power estimates that depend on simplifying assumptions such as a lack of covariate effects, or on external ICC estimates.

Our purpose in the current article is to propose a means of avoiding these problems and obtaining the greatest possible verisimilitude in power calculation. In the Methods section, we discuss the general approach to power assessment using resampling methods, and outline two distinct settings in which they are likely to be useful. In the Results section we describe an application in which we implemented the method, and show the resulting power assessment.

Methods

In short, we propose to use bootstrapped samples to assess statistical power, modifying the samples as necessary to generate the desired alternative. This approach is a rather natural one. Bootstrapping for power calculation has been described previously in a few specific applications, but its generality, flexibility, benefits, and heuristic motivation have not been fully explored, to the best of our knowledge [23-26]. Nor has the application or its unique advantages in cluster randomized trials been described.

Bootstrap resampling is simply sampling from observed data, with replacement. Heuristically, the idea is that the observed data represent the population from which they were drawn, and thus sampling an observation from among the observed data can be substituted for sampling from the population itself.

For estimating the power in a typical medical study, the method requires having relatively detailed data before the power must be calculated. Ideally, this data should be as similar to the prospective study data as possible; one example would be baseline data which would also then be used in the study itself. Another setting where the method may be possible is in laboratory studies, where new experiments may be quite similar to completed experiments. We will describe how the approach might be implemented in each of these cases.

We begin with the laboratory experiment, a simple non-clustered setting, to introduce the idea. Suppose conditions "A" and "B" were compared in "Study I", which has been completed. Now we wish to assess the power for a new experiment, "Study II", where we will compare condition A to condition, "C", a modification of condition B. Let us assess the power under the alternative that the mean of condition C in Study II is 5 units greater than was observed for condition B in study I. We denote the data observed under condition A in Study I as $x_{A1}, \ldots, x_{An_1}$ and the data observed under condition B in Study I as $x_{B1}, \ldots, x_{Bn_2}$. Our bootstrap power calculation for Study II proceeds as follows:

1. Sample $n_1$ values from $x_{A1}, \ldots, x_{An_1}$ using simple random sampling with replacement; denote these values $x_{A1}^*, \ldots, x_{An_1}^*$
2. Sample $n_2$ values from $x_{B1}, \ldots, x_{Bn_2}$ using simple random sampling with replacement; denote these values $x_{B1}^*, \ldots, x_{Bn_1}^*$
3. Add 5 to each of the values from step 2 and denote the values thus modified as $x_{C1}^*, \ldots, x_{Cn_2}^*$; this is how we include the alternative in the calculation
4. Perform the test comparing $x_{A1}^*, \ldots, x_{An_1}^*$ to $x_{C1}^*, \ldots, x_{Cn_2}^*$, record whether the null was rejected or not
5. Repeat steps 1-4 many times
6. The proportion of rejections is the estimated power

A diagram of steps 1 through 4 is presented in Figure 1.

The value of this approach is immediately obvious. Suppose the distribution in the second group is exponential, while that of the first is normal. An analytic approach to the power that accurately incorporates this difference in outcome distribution is not likely to be available. The choice of test with such distributions might be non-trivial, but the above routine will quickly generate the estimated power regardless of the chosen test; the algorithm above does not even specify a test. If we assume the new second condition will change the scale of the outcome, instead of or in addition to the location, we could easily modify step 3 in the above algorithm, and still generate the desired result.

Next, let us consider a cluster-randomized trial with a baseline observation period. Suppose we have collected baseline data on the presence or absence of an outcome, among individuals at several sites or clusters. We might be able to do this before the study was fully funded by using electronic medical records, for example. Each site may have a different number of subjects. We plan to use the collected data as a baseline against which we will compare data collected on other subjects while an intervention is applied to a random subset of sites. Suppose we need to know how much power we would have, given the intervention increases the odds of an outcome at each site by a factor of 2.

Denote each subject seen at cluster $C$ in the baseline period as $s_{Cj}$, $j = 1, \ldots, n_C$, and the outcome for subject $s_{Cj}$ as $y_{Cj} = 1$ if the outcome is observed and 0 otherwise. Our bootstrap power calculation would proceed along the following steps:

1. Within each cluster, resample $n_C$ observations, call these bootstrapped subjects $s_C^B$. These will serve as the baseline data.
2. Randomize clusters to the control or intervention condition using whatever randomization strategy is planned for the actual study; this step may involve stratification by features of the "observed" $s_C^B$.
3. Again, within each cluster, resample $n_C$ observations, call these subjects $s_C^I$. These will serve as the intervention period data. There will be new individuals in the second period; this is why we resample again.
4. For clusters assigned to the intervention condition, calculate the modified within-cluster probability in the intervention period: Calculate the proportion where $y_{Cj}^I = 1$, denote this proportion $p_C^B$, and from this calculate the odds within the cluster. Multiply by 2, and solve the odds for the probability implied. Call this $p_C^I$.
5. For clusters assigned to the intervention condition, set all $y_{Cj}^I = 0$, and set them to have $y_{Cj}^I = 1$ with probability $p_C^I$. Optionally, a more sophisticated approach would be to calculate the difference $p_C^D = p_C^I - p_C^B$; this is the additional probability of the outcome in cluster $C$ in the intervention period. Under this approach we would retain the original values of $y_{Cj}^I$. Then, among subjects with $y_{Cj}^I = 0$, we would then change their outcome value so that $y_{Cj}^I = 1$ with probability $n_C p_C^D / n_C(1 - p_C^B)$.

6. Perform the planned analysis (say, a generalized linear mixed logistic regression model) on the data set comprising the $s_C^B$ and the modified $s_C^I$ from step 5; record whether the null hypothesis was rejected or not.
7. Repeat steps 1-6 many times.
8. The proportion of rejections is the power.

A diagram of steps 1 through 6 is presented in Figure 2.

Again, the advantages are obvious. There is no concern about how to incorporate the variable cluster size into the design effect and analytic approaches. There is no need to estimate the ICC or $\sigma_b^2$, which would be necessary for other approaches. Similarly, we need not estimate the precision of our estimated ICC or $\sigma_b^2$. Though it might be possible to estimate the ICC or $\sigma_b^2$ using the baseline data and then proceed with a simulation or analytic approach, it might be awkward to incorporate the variability of the estimate.

Note that the resampling is within cluster, so that the correlation within cluster is maintained, and so that independence across the bootstrapped items is also possible. The observed distribution of cluster sizes is also quite naturally maintained. Also note that the randomization to study condition is placed within the power assessment process so that it can depend on the values in the resampled baseline data. The somewhat complex optional formulation of step 5 is intended to retain the actual observations where the outcome is observed, and only add to them. This would possibly facilitate the incorporation of covariates.

Results

Here we present a real application. In 2012, we received a planning grant from the United States National Institutes of Health (National Institutes of Health Health Care Systems Research Collaboratory - Pragmatic Clinical Trials Demonstration Projects) to plan a trial of decolonization to reduce clinical cultures of certain drug-resistant organisms and bloodstream infections in hospitals. The intervention was to involve daily bathing of all patients with an antiseptic soap, plus a nasal antibiotic ointment for patients who harbored antibiotic-resistant bacteria. For cost and practicality reasons, this intervention was to be implemented at the hospital level, rather than at the unit or patient level. Thus, a cluster-randomized trial was planned. Data collection would use the hospitals' routine electronic records, so a baseline observation period was feasible.

During the planning year, we recruited 55 hospitals and collected four months of data. We were then invited to pursue funding to actually perform the trial. That trial has since been funded and is ongoing as of this writing. Further details are available from clintrials.gov, identifier NCT02063867.

Due to variable length-of-stay in the hospital, we will treat the outcomes as time-to-event or survival data. We plan to use a proportional hazards model, or Cox model, to assess the effectiveness of the intervention, with frailties to account for randomization by cluster [27-29]. We performed a version of the bootstrap power calculation described above. We know of no analytic approach useful in this setting. The primary outcome is time to clinical culture with methicillin-resistant *Staphylococcus aureus* or vancomycin-resistant *Enterococcus*. These are both important antibiotic-resistant bacteria. Secondary outcomes include time until clinical culture with a gram negative multi-drug resistant organism and time until bacteremia resulting from any pathogen. Fortunately these organisms and infections are currently rare, and the event rate per 1,000 attributable days for them respectively, is 2.2, 0.6, and 1.1.

The consensus among study planners was that we should assess power assuming the rate of events would be 20% smaller with the intervention than in the baseline. Our bootstrap power routine resembles that described above for a dichotomous outcome. As before, denote each subject seen at cluster C in the baseline as $s_{Cj}, j = 1, \ldots, n_C$. In the baseline period, we observed the time of event for subject $s_{Cj}$, which we denote $t_{Cj}$, and the censoring indicator $c_{Cj} = 1$ if the event is the outcome of interest and 0 if it is censored, which includes discharge from the hospital with no infection, for example. We also observed the time of discharge for patients who actually were infected. We'll denote this time as $t'_{Cj}$. Our routine looks something like this:

1. Within each cluster, resample $n_C$ observations. Call these subjects $s_C^B$. These will serve as the baseline data in our assessment.
2. Randomize clusters to the control or intervention condition using the randomization strategy we plan to employ in the actual study.
3. Again, within each cluster, resample $n_C$ observations, call these subjects $s_C^I$. These will serve as the intervention period data in our power assessment. In this design, there are new individuals in the second period; this is why we resample again instead of re-using the $s_C^B$.
4. For clusters assigned to the intervention condition, randomly reassign 20% of the resampled event cases (where $c_{Cj} = 1$) to instead have $c_{Cj} = 0$. For these subjects, replace $t_{Cj}$ with $t'_{Cj}$. This replaces the observed event time with the discharge time and changes the event indicator to indicate censoring.
5. Fit the frailty model to the data set comprising the $s_C^B$ and the modified $s_C^I$, record whether the null hypothesis was rejected or not.
6. Repeat steps 1-5 many times.
7. The proportion of rejections is the power.

An example using simulated data in SAS is shown in Additional File 1.

We note that the substitution of $t'_{cj}$ for $t_{cj}$, replacing the event time with the discharge time when an event is removed, is not perfect. Ideally, we would substitute the date of discharge which would have occurred had there been no event, but of course this is unknowable. The date of discharge after an event might well be later than this unknowable value, since the event itself may delay the time of discharge, for example. It is possible that a better choice would be to leave the time of event unchanged, while changing it to a censoring rather than event time, effectively discharging the patient at the event time, although this would almost certainly censor non-event time. In our application, however, the event rate is very small, so the difference between these imperfect choices is unlikely to be meaningful.

There was a further complication. The baseline data in hand included 4 months of recruitment, but the planned total baseline accrual period was 12 months. The intervention period was scheduled for 18 months. We resolved this issue by referring back to the heuristic behind the bootstrap: the observed sample represents the original population. Why not bootstrap more than $n_c$ samples from among the $n_c$ observed subjects in each cluster in the baseline period? In traditional uses of the bootstrap, for example to obtain confidence limits for statistics which have difficult asymptotic properties, this idea would lead to biased results—narrower confidence limits than appropriate. But in our setting this argument is not relevant. We sampled $3 * n_c$ observations from each cluster for the baseline in step 1 and $4.5 * n_c$ from each cluster in step 3.

Results are shown in Table 1. Power for the primary outcome and two selected secondary outcomes were assessed; we repeated the above process for each of these. We also considered four values of the intervention effect by varying the percent selected to have their events removed in step 4 above.

The results show that for the primary outcome, there is ample power to detect the anticipated effect of preventing 20% of events. There is notably less power for the secondary outcomes. Note that we also present an effect of 0%. A 0% effect is implemented by not altering the outcomes for any subjects, in step 4. In this case the null hypothesis is true, which may strike some readers as meaning we are assessing something other than power. We include this effect as a face validity check to ensure that the power assessment process and implementation are correct: the probability of rejecting the null hypothesis must be the same as our rejection level—5% in this case. The results show that the face validity test was passed.

As is demonstrated by the table, the results of the bootstrap power calculation, as with simulation-based power procedures, are estimates. The result of the calculation is a proportion, and the precision of the estimate, controlled by the number of bootstrap cycles in step 6, should be reflected by providing confidence limits based on the properties of the binomial distribution. Here we use exact limits.

Discussion

We describe the application of bootstrap resampling to the problem of power calculation. Like the bootstrap itself, the approach is very general. It can be used in laboratory settings and in any application where fairly extensive preliminary data can be obtained before power calculations are necessary. While the simple application has been proposed previously, we described its application and explored the nuances of use in the setting of cluster-randomized trials [23-26]. We also showed a real example of estimating power for a cluster-randomized trial for infection prevention in hospitals. The application to time-to-event outcomes allows power estimation for a setting in which analytic results are not available.

The bootstrap power approach offers several advantages beyond its ability to account for arbitrary complexity in the structure of the data and the fact that it does not rely on estimates from the literature. Primary among these is that it can use the precise analysis method contemplated for the planned study, an advantage shared with simulation approaches, but without the requirement of verisimilitude in the simulation. For example, covariates can easily be incorporated without making assumptions about their joint distribution. It also incorporates variability in key parameters without further consideration of the analyst. Another type of advantage is the ease of implementing the thinking of study planners. For example, in the time-to-event application, the expert consensus was that 20% of the infections might be prevented by the intervention. Using the bootstrap power approach, we were able to implement that effect directly, without having to consider implications for model parameters such as the hazard ratio.

Additionally, while we demonstrated simple alternative hypotheses, it would be trivial to implement complex ones. For example, we could change the mean in the laboratory experiment by the same amount by leaving half the experimental subjects unchanged and doubling the effect in the other half. In our real example, we could change the shape of the survival curve by removing events preferentially among early events. Another advantage is that it is entirely generic—any analytic method can be inserted into the data analysis step, and the power assessment algorithm will be unchanged. This suggests that bootstrap power calculations could easily be used to compare the power of two competing analyses in a particular data setting.

The primary weakness of the approach is the reliance on the availability of detailed data. We provide the example of laboratory experiments as a case where detailed data may well be available, and demonstrate a real example of a large human trial in which it is possible. And we believe it will frequently be possible in pragmatic projects using established care delivery systems, where detailed data is often easily available through existing electronic records. On the other hand, in such settings, it may also be possible to estimate key parameters, such as the ICC, from the preliminary data, diminishing the advantage of the bootstrap. We believe the bootstrap power approach is still markedly superior, however, in that it incorporates the variability of these parameters, as compared with fixed estimates, or, at best, large variability for estimates of second moments.

A few other items bear some attention. One is the particular utility of being able to bootstrap more than N subjects. This allowed us to expand the time scale of the available baseline in the cluster-randomized trial application. It also suggests that we can assess the number of subjects needed to achieve a given power, as is often desirable. Another is the unique feature that there is no need to explicitly estimate parameters from the collected data. Thus for the cluster-randomized trial example, we did not calculate the baseline rate or survival curve for each event, or need to know the sample size available at each cluster. These are features that play heavily into analytic and simulation-based power assessments. Finally, as noted above, power estimates from bootstrap and simulation methods are explicitly estimates, and can and should be accompanied by confidence limits. Ironically, power calculations from analytic methods treat their inputs as fixed and offer no formal means of assessing uncertainty. At best, we may vary parameters informally to demonstrate the effects of uncertain inputs to the formulae.

Conclusions

Power calculation by bootstrap is the simple proposal to use resampling techniques to generate data under the alternative hypothesis and to use replication to assess power under that hypothesis. Bootstrap power calculation is a powerful tool that offers unique advantages in several research settings. It allows power to be driven by detailed baseline data and avoids weaknesses common to other approaches to power, including the need to assume literature-based estimates apply to the population under study and the need to find viable estimates of all parameters in the analysis. It should be particularly useful, as demonstrated, in application to cluster-randomized trials.

List of abbreviations

ICC: Intracluster Correlation Coefficient

Competing interests

The authors declare that they have no competing interests.

Authors' contributions

KK generated the idea for the method and drafted the manuscript. SH edited the manuscript for important intellectual content. Both authors have given final approval.

Acknowledgements

We appreciate the assistance of Taliser Avery in implementing the proposed method as presented in the results. Each author's effort on the manuscript was supported by NIH grant 1UH2AT007769-01.

27. Cox DR. **Regression Models and Life Tables.** *Journal of the Royal Statistical Society, Series B* 1972, 20: 187–220

28. Ripatti S, and Palmgren J. **Estimation of Multivariate Frailty Models Using Penalized Partial Likelihood.** *Biometrics* 2000, 56:1016–1022

29. Hayes RJ, Moulton LH: *Cluster Randomized Trials.* Boca Raton, Chapman&Hall/CRC 2009

Figures

Figure 1: Diagram for bootstrap power calculation in laboratory experiment

Figure 2: Diagram for bootstrap power calculation in a cluster-randomized trial with a baseline observation period

**Table 1: Power and exact 95% CI for power for primary and select secondary outcomes***

| Intervention Effect | MRSA or VRE clinical cultures | Gram negative Multi-drug resistant clinical cultures | All-pathogen bacteremia |
| --- | --- | --- | --- |
| 0% | 5.6% (4.3-7.2%) | 4.1% (3.0 – 5.5%) | 5.2% (3.9 – 6.8%) |
| 10% | 35% (32-38%) | 14% (12 – 16%) | 22% (19 - 24%) |
| 20% | 93% (91-95%) | 44% (41 – 47%) | 67% (63 - 69%) |
| 30% | 100% (99.6-100%) | 83% (81 – 86%) | 97% (96 – 98%) |

*Based on 1000 bootstrap samples for each effect size and outcome.

Additional Files

Additional File 1 is a .txt file with SAS code demonstrating the method.

Study I

Condition A: $x_{A1}, \ldots, x_{An_1}$

Condition B: $x_{B1}, \ldots, x_{Bn_2}$

Step 1: bootstrap → $x^*_{A1}, \ldots, x^*_{An_1}$

Step 2: bootstrap → $x^*_{B1}, \ldots, x^*_{Bn_1}$

Step 3: Add 5 → $x^*_{C1}, \ldots, x^*_{Cn_2}$

Step 4: Test $x^*_{A1}, \ldots, x^*_{An_1}$ vs. $x^*_{C1}, \ldots, x^*_{Cn_2}$

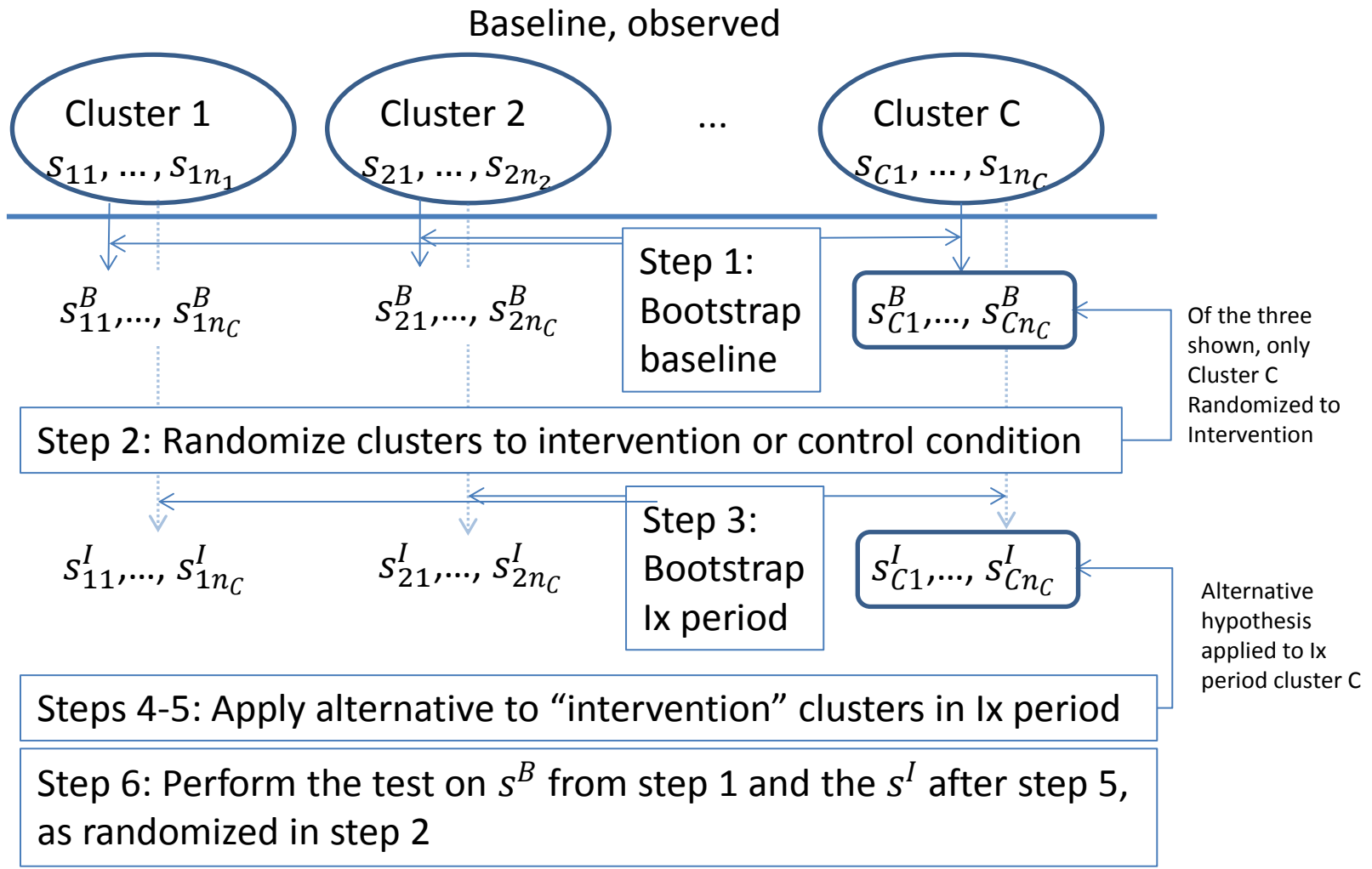

Figure 2: Diagram for bootstrap power calculation in condition*period cluster-randomized trial

```sas
/* Demonstration code to do bootstrap power, for frailty models */

/* First, generate some "observed" baseline data that we will resample
from.
   This will be replaced with real data in practice. */

/* This simulation is not meant to be especially accurate! */
data simfrail;
beta1 = 2;
beta2 = -1;
do hosp = 1 to 60; *frailty loop;
   frailty = normal(0) * sqrt(.25);
/* as and additional covariate, assume 1 ward of each of 4 types at each
hosp */
     do wardtype = 1 to 4;
/* id = patient */
     do id = 1 to ceil(33 + (33*uniform(0)));
          *add variability to ward size: ward size  = 33 - 65;
       x2 = (normal(0) gt 0); * covariate;
         mean = exp(1.5 - log(2)*x2 + frailty);  * mean event time;
       event = rand("EXPONENTIAL") * mean;     * observed event time;
         inelig = event + 2;                    * time of censoring--
                      recall that this is not an accurate
simulation!;
       censored = (uniform(0) gt (.1 + (.015 * x2)));
         * indicator of censoring: ~89% censored ;
       if censored then time = inelig;
           else time = event;    * implements censoring;
     output;
       end;
     end;
  end;
run;

/****************************************/
/* bootstrap                            */
/****************************************/
proc sort data = simfrail; by hosp wardtype; run;

%let nreps = 100;  * number of iterations of the process;

/* bootstrap for "baseline" */
proc surveyselect data=simfrail noprint method = urs samprate = 1
  out=base outhits reps= &nreps; /* reps = number of bootstrapped data
sets */
strata hosp wardtype;   /* observed n/ward maintained */
run;
```

```sas
/* bootstrap for "Ix period" */
/* sample subjects as above */
proc surveyselect data=simfrail noprint method = urs samprate = 1
  out=ix outhits reps= &nreps; /* reps = number of bootstrapped data sets
*/
strata hosp wardtype;  /* observed n/ward maintained */
run;

/* Add both periods together */
/* add period indicator */
data c1;
set
base (drop = numberhits expectedhits samplingweight in = base)
ix (drop = numberhits  expectedhits samplingweight  in = ix);
if base then period = 0;
else if ix then period = 1;
run;

proc sort data = c1; by replicate hosp; run;

/****************************************/
/* randomize hosps to arms              */
/****************************************/
/* first, order on hosp size, within replicate */
proc summary data = c1;
class replicate hosp;
var censored;
output out=c1a n=nhosp mean=pct_censored;
  /* pct_censored = 1 - rate of outcome.  Matching on one is the same as
matching on the other */
run;

/* _freq_ contains the number of obs/hosp; type < 3
            has various summaries for the data set */
proc sort data=c1a (where = (_type_ eq 3)) out=c1b; by replicate nhosp;
run;

/* now, make strata of 4 hosps by size; within these, rank by percent
censored */
data c1c;
set c1b;
strata = int((_n_ -1)/4);
run;

proc sort data = c1c; by replicate strata pct_censored; run;
```

```sas
/* now, randomize, in pairs */
data c1d;
set c1c;
retain arm;
if int((_n_ -1)/2) = (_n_ -1)/2 then arm = uniform(0) > .5;
  else arm = 1 - arm;
run;

proc sort data = c1d; by replicate hosp; run;

/* now, merge arm status into data */
/* also, implement alternative truth */
data c2;
merge c1 c1d (keep = replicate hosp arm);
by replicate hosp;
if arm eq 1 and censored eq 0 and period eq 1 then do;
  censored = (uniform(0) lt .2);  /* alternative risk reduction percent in here */
  if censored eq 1 then time = inelig;
  end;
run;

proc sort data = c2; by replicate hosp; run;

/*****************************************/
/* fit the model to each replicate       */
/*****************************************/
ods select none;
ods output type3 = kktype3 parameterestimates = kkpe;
proc phreg data=c2;
by replicate;
  class hosp x2(ref='0') arm(ref='0') period(ref='0');
  model time*censored(1) = x2 wardtype arm|period;
  random hosp / noclprint;
run;
ods select all;

/* check to see when rejected */
data kkpesum; set kkpe (where = (parameter="arm*period"));
reject = (probchisq < .05);
run;

/* generate proportion rejected == power, plus CI */
proc freq data = kkpesum;
tables reject / binomial(level='1');
run;
```

```
/* results, will vary with random seed in data generation and proc
surveyselect:
89% power, CI 81-94%.

                                The FREQ Procedure

                                                      Cumulative
Cumulative
                    reject     Frequency      Percent     Frequency
Percent
ƒƒƒƒƒƒƒƒƒƒƒƒƒƒƒƒƒƒƒƒƒƒƒƒƒƒƒƒƒƒƒƒƒƒƒƒƒƒƒƒƒƒƒƒƒƒƒƒƒƒƒƒƒƒ
                         0         11        11.00            11
11.00
                         1         89        89.00           100
100.00

                              Binomial Proportion
                                 for reject = 1
                         ƒƒƒƒƒƒƒƒƒƒƒƒƒƒƒƒƒƒƒƒƒƒƒƒƒƒƒƒƒƒƒƒƒ
                         Proportion                0.8900
                         ASE                       0.0313
                         95% Lower Conf Limit      0.8287
                         95% Upper Conf Limit      0.9513

                         Exact Conf Limits
                         95% Lower Conf Limit      0.8117
                         95% Upper Conf Limit      0.9438

                           Test of H0: Proportion = 0.5

                         ASE under H0              0.0500
                         Z                         7.8000
                         One-sided Pr >  Z         <.0001
                         Two-sided Pr > |Z|        <.0001

                              Sample Size = 100

*/
```